\documentclass[preprint]{aastex63}

\usepackage[T1]{fontenc}

\usepackage{bm}        
\usepackage{amssymb, amsmath}   
\usepackage{cancel}
\usepackage{color}

\newcommand\aastex{AAS\TeX}

\DeclareMathAlphabet{\mathsfit}{\encodingdefault}{\sfdefault}{m}{sl}
\SetMathAlphabet{\mathsfit}{bold}{\encodingdefault}{\sfdefault}{bx}{sl}
\DeclareMathOperator\erf{erf}
\newcommand{\vect}[1]{\bm{#1}}

\DeclareMathOperator{\sech}{sech}
\newcommand{\tensorsym}[1]{\bm{\mathsfit{#1}}}

\received{****}
\revised{****}
\accepted{****}
\submitjournal{ApJ}

\shorttitle{\aastex\ The Thermal Effects on Collisionless Magnetic Reconnection Rate}
\shortauthors{Li et al.}

\begin{document}

\title{The Effect of Thermal Pressure on Collisionless Magnetic Reconnection Rate}

\correspondingauthor{Xiaocan Li}
\email{Xiaocan.Li@dartmouth.edu}

\author[0000-0001-5278-8029]{Xiaocan Li}
\affil{Dartmouth College, Hanover, NH 03750 USA}

\author{Yi-Hsin Liu}
\affiliation{Dartmouth College, Hanover, NH 03750 USA}

\begin{abstract}
  Modeling collisionless magnetic reconnection rate is an outstanding challenge in basic plasma physics research. While the seemingly universal rate of an order  $\mathcal{O}(0.1)$ is often reported in the low-$\beta$ regime, it is not clear how reconnection rate scales with a higher plasma $\beta$. Due to the complexity of the pressure tensor, the available reconnection rate model is limited to the low plasma-$\beta$ regime, where the thermal pressure is arguably negligible. However, the thermal pressure effect becomes important when $\beta \gtrsim \mathcal{O}(1)$. Using first-principle kinetic simulations, we show that both the reconnection rate and outflow speed drop as $\beta$ gets larger. A simple analytical framework is derived to take account of the self-generated pressure anisotropy and pressure gradient in the force-balance around the diffusion region, explaining the varying trend of key quantities and reconnection rates in these simulations with different $\beta$. The predicted scaling of the normalized reconnection rate is $\simeq \mathcal{O}(0.1/\sqrt{\beta_{i0}})$ in the high $\beta$ limit, where $\beta_{i0}$ is the ion $\beta$ of the inflow plasma.
\end{abstract}


\section{Introduction}

Magnetic reconnection is a ubiquitous fundamental plasma process that reorganizes the magnetic topology and releases the magnetic energy into plasma kinetic energies~\citep{Zweibel2009Magnetic}. It occurs in laboratory experiments~\citep{Yamada2006}, confined fusion devices~\citep{Yamada1994} and drives explosive magnetic energy release in space~\citep{Schindler1974,Oieroset2002Evidence}, solar~\citep{Masuda1994Loop, Lin2011Energy}, and astrophysical plasmas~\citep{Colgate2001Origin,Zhang2011Internal}. A long-standing problem in reconnection studies is how fast reconnection processes available magnetic flux, the so-called reconnection rate problem~\citep[see][and reference therein]{Cassak2017JPP}.

Many reconnection models have been constructed to explain the reconnection rate observed---about 0.1 in normalized units---in numerical simulations~\citep[e.g.,][]{Birn2001Geospace} and space~\citep[e.g.,][]{Wang2015Dep} and solar plasmas~\citep[e.g.,][]{Yokoyama2001,Qiu2002}. The most famous one is the Sweet-Parker model~\citep{Sweet1958,Parker1957,Parker1963}, which is the first quantitative reconnection model but predicts a rate far too slow to explain, for example, solar flares. In this model, the current sheet is long and thin, limiting the inflow flux of plasmas and, therefore, the reconnection rate is low. The following Petschek model~\citep{Petschek64} predicts a much shorter current sheet and a much higher reconnection rate. However, numerical simulations have demonstrated that it requires an {\it ad hoc} localized resistivity to be stable, and thus the origin of the localization is not captured in this model~\citep{Biskamp1986,Sato1979}. Recent progress features collisionless physics in the diffusion region to be the key to produce fast reconnection, notably Hall physics~\citep[e.g.,][]{Shay2001} or secondary islands~\citep{Daughton2007,Liu2014}. The up-to-date model by~\citet{Liu2017Why} shows that the value of fast reconnection rate is insensitive to these diffusion-region-scale physics but is instead constrained by the mesoscale magnetic geometry and force balance. This model expresses the reconnection rate as a function of the exhaust opening angle; it predicts the fast rate on the order of $\mathcal{O}(0.1)$ persists for a wide range of opening angles,  and the maximum plausible reconnection rate is bounded by $\simeq 0.2$ in the low-$\beta$ regime.

The thermal pressure is often neglected in those models, which might be valid in the low-$\beta$ regime and suitable for studying reconnection in solar flares, Earth's magnetotail, or magnetically dominated astrophysical plasmas. However, thermal pressure could dominate the plasma dynamics with a higher $\beta\equiv P/(B^2/8\pi)$, as in the outer heliosphere~\citep[$\beta$ up to 10,][]{Drake2010Magnetic,Schoeffler2011}, in the hot intracluster medium (ICM) of galaxy clusters~\citep[$\beta\sim10^{2-4}$,][]{Carilli2002,Schekochihin2006}, or at the Galactic center~\citep[$\beta\sim10^{1-2}$,][]{Marrone2007}.  Self-generated pressure anisotropy and/or pressure gradient upstream~\citep{Egedal2013Review} and downstream ~\citep{Liu2011,Liu2012,Haggerty2018,Bessho2010Fast} of the diffusion region could affect the force-balance and reduce the outflow speed. Therefore, it is critical to include the thermal correction to a reconnection model for studying the reconnection rate in the high-$\beta$ regime.

In this paper, we extend the reconnection model by~\citet{Liu2017Why} to include the thermal correction. By including the pressure anisotropy and pressure gradient force in the inflow and outflow force balance equations, we get the magnetic field immediately upstream of the diffusion region and the outflow speed as a function of both the plasma $\beta$ and the exhaust opening angle. A prediction of the normalized reconnection rate in a given $\beta$ can be obtained by maximizing the rate with respect to the opening angle, which scales as $\simeq \mathcal{O}(0.1/\sqrt{\beta_{i0}})$ in the high-$\beta$ limit, where $\beta_{i0}\equiv 8\pi P_{i0}/B_{x0}^2$ defines the ratio of the inflow ion thermal pressure ($P_{i0}$) and the magnetic pressure of the asymptotic magnetic field $B_{x0}$. In Section~\ref{sec:num}, we perform 2D kinetic simulations with different plasma $\beta$ to show the scaling of the reconnection rate and related quantities with plasma $\beta$. In Section~\ref{sec:rate_model}, we present the extended model and discuss its predictions in the low-$\beta$ and high-$\beta$ limits. In Section~\ref{sec:compare}, we compare the model results with the simulations. In Section~\ref{sec:con}, we discuss the conclusions and implications based on our results.

\section{Numerical Simulations}
\label{sec:num}

\begin{figure}[ht!]
  \centering
  \includegraphics[width=0.5\textwidth]{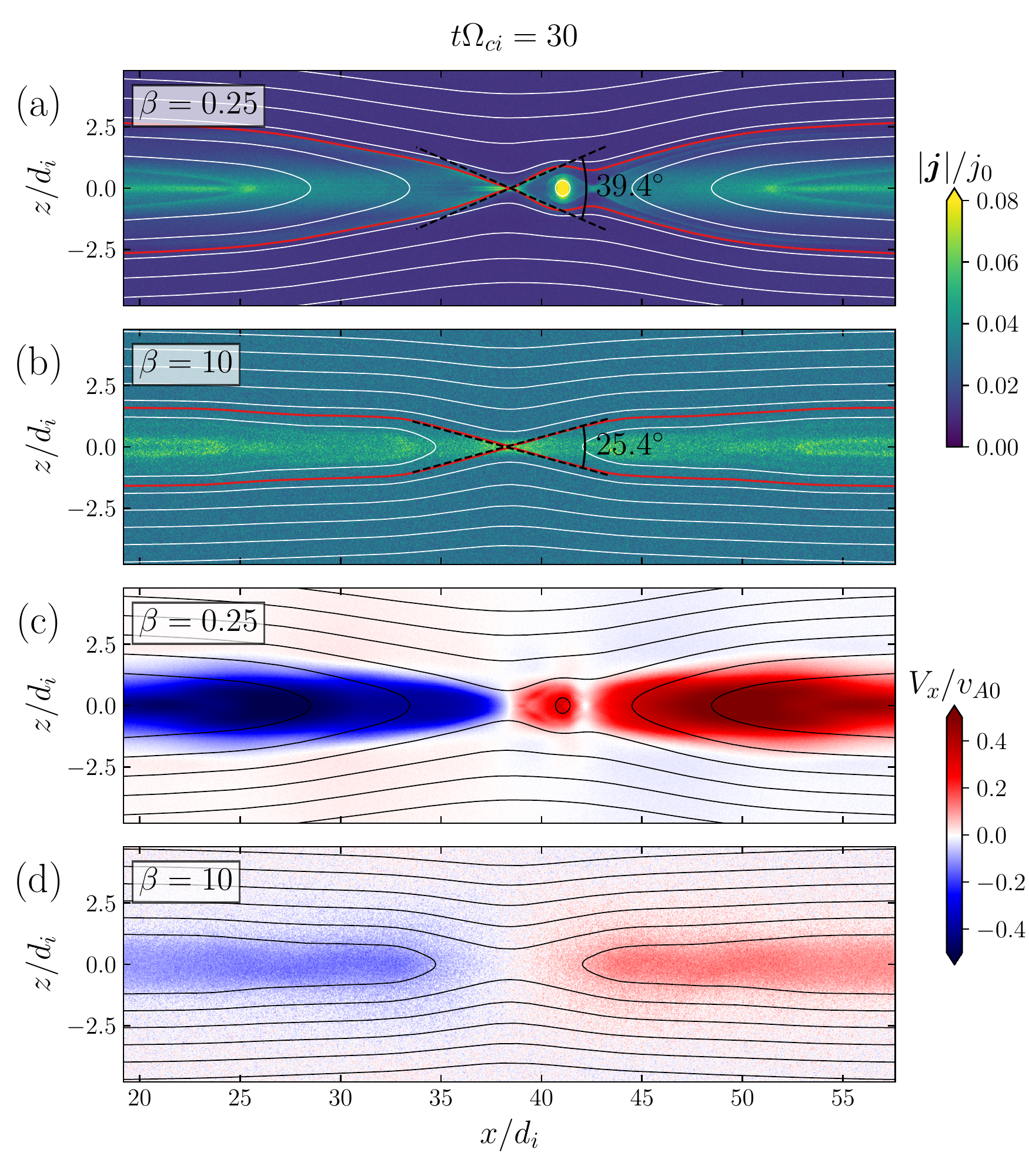}
  \caption{\label{fig:absj_vout}
    Reconnection layer near the main X-point. The top two panels show the current density for the runs with (a) $\beta=0.25$ and (b) $\beta=10$. The red lines indicate the separatrix. The point where the top line meets the bottom one indicates the X-point. Assuming its coordinate is $(x_0, z_0)$, we then calculate the angle $\theta\equiv\arctan(|z-z_0|/|x-x_0|)$ for all the points $(x, z)$ along the separatrix and get the maximum values for the four branches staring from the X-point ($\theta_1$--$\theta_4$). The four maximums are similar except when a plasmoid is ejected and opens up the right side of the exhaust in the run with $\beta=0.25$ (panel (a)). We evaluate  $\theta_0=\text{mean}(\theta_i)$. The exhaust opening angle is then about $2\theta_0$. The dashed lines with a slope $\pm\tan\theta_0$ indicate the boundaries for obtaining the opening angles. The bottom two panels show the outflow velocity $V_{ix}$ normalized by the upstream Alfv\'en speed $v_{A0}$ (which is kept the same for all cases).
  }
\end{figure}

We carry out 2D kinetic simulations of magnetic reconnection in plasmas with $\beta=0.25, 1, 10, 40$, studying how the reconnection rate responds to this change. The simulations were performed using the VPIC particle-in-cell code~\citep{Bowers2008PoP}, which solves Maxwell's equations and the relativistic Vlasov equation. The simulation employs a Harris current sheet with the magnetic profile $\vect{B}=B_{x0}\tanh(z/\lambda)\hat{x}$, where $B_{x0}$ is the reconnecting magnetic field and $\lambda$ is the half-thickness of the current sheet. The simulation size is $L_x\times L_z = 76.8d_i\times76.8d_i$ that spans domain $[0,L_x] \times [-L_z/2,L_z/2]$, where $d_i$ is the ion inertial length. We chose a proton-to-electron mass ratio $m_i/m_e=400$. The plasma consists of a Harris sheet component with a peak density $n_0$ and a background component with a uniform density $n_b$. Its density profile satisfies $n=n_0 \sech^2(z/\lambda)+n_b$ to maintain the initial pressure balance. In all runs, we choose $\omega_{pe}/\Omega_{ce}=2$, where the plasma frequency $\omega_{pe}=(4\pi n_0 e^2/m_e)^{1/2}$ and the electron gyro-frequency $\Omega_{ce}=e B_{x0}/m_e c$, resulting in an Alfv\'en speed $v_{A0}\equiv B_{x0}/(4\pi n_0m_i)^{1/2}=0.025c$. To have a similar dynamical time ($\sim L_x/v_{Ab}$, where $v_{Ab}\equiv B_{x0}/(4\pi n_bm_i)^{1/2}$) for different runs, we choose $n_b=n_0$ in all runs to fix $v_{Ab}=v_{A0}$. Electrons and ions have the same uniform temperature $T_s$ in the sheet component. The initial pressure balance $2n_0kT_s=B_{x0}^2/8\pi$ results in an electron thermal speed $v_\text{the}\equiv(kT_s/m_e)^{1/2}=(v_{A0}/2)(m_i/m_e)^{1/2}=0.25c$ for the sheet component. Electrons and ions have the same uniform temperature $T_0$ in the background component. In the four runs, $T_0=0.25T_s$, $T_s$, $10T_s$, and $40T_s$, resulting in a plasma $\beta_0\equiv16\pi n_bkT_0/B_{x0}^2=0.25$, 1, 10, and 40, respectively. Note that for this paper we will use $\beta_0$ and $\beta$ interchangeably when it does not cause confusion. Since electrons and ions have the same temperature, $\beta_{e0}=\beta_{i0}=\beta_0/2$ in this study. The grid numbers are $n_x\times n_z = 12288\times12288$ for the run with $\beta=0.25$ and $6144\times6144$ for runs with higher plasma $\beta$. For electric and magnetic fields, we employ periodic boundaries along the $x$-direction and perfectly conducting boundaries along the $z$-direction. For particles, we employ periodic boundaries along the $x$-direction and reflecting boundaries along the $z$-direction. A localized initial perturbation is added to induce reconnection with a single X-line. The $y$-component of the vector potential of the perturbation is
\begin{align}
  \delta A_y = -\frac{\delta B}{(2\pi/L_p)^2+(1/L_c^2)}
  \left[\frac{1}{L_c}\sin\left(\frac{2\pi|x'|}{L_p}\right)
  +\frac{2\pi}{L_p}\cos\left(\frac{2\pi x'}{L_p}\right)\right]
  \exp\left(-\frac{|x'|}{L_c}\right)\cos\left(\frac{\pi z}{L_z}\right),
\end{align}
where $x'=x-0.5L_x$, $\delta B$ controls the amplitude of the perturbation, $L_p$ is the length scale of the sinusoidal perturbation along the $x$-direction, and $L_c$ is the length scale that controls how fast the perturbation decays along the $x$-direction. We have chosen $\delta B=0.165B_{x0}$, $L_p=L_x/12$, and $L_c=L_x$ in our simulations.

Figure~\ref{fig:absj_vout} shows the current layer near the primary X-line for runs with $\beta$ = 0.25 and 10. One single X-line forms in the reconnection layer under local perturbation, and the reconnection exhaust gradually opens up. The exhaust's opening angle is between 20$^\circ$ and 30$^\circ$ and can get over \text{40}$^\circ$ when a magnetic island forms in the case with $\beta=0.25$. Fig.~\ref{fig:absj_vout}(c) shows that the reconnection outflow $V_x$ can reach $0.5v_{A0}$ in the $\beta=0.25$ case, while Fig.~\ref{fig:absj_vout}(d) shows $V_x$ is below $0.2v_{A0}$ in the $\beta=10$ case. Given a similar exhaust opening angle, the reduction of the reconnection outflow at the high-$\beta$ regime implies a weaker outflow motional electric field that potentially leads to a lower reconnection rate. We will model this reconnection outflow reduction in the next section.

\begin{figure}[ht!]
  \centering
  \includegraphics[width=0.6\textwidth]{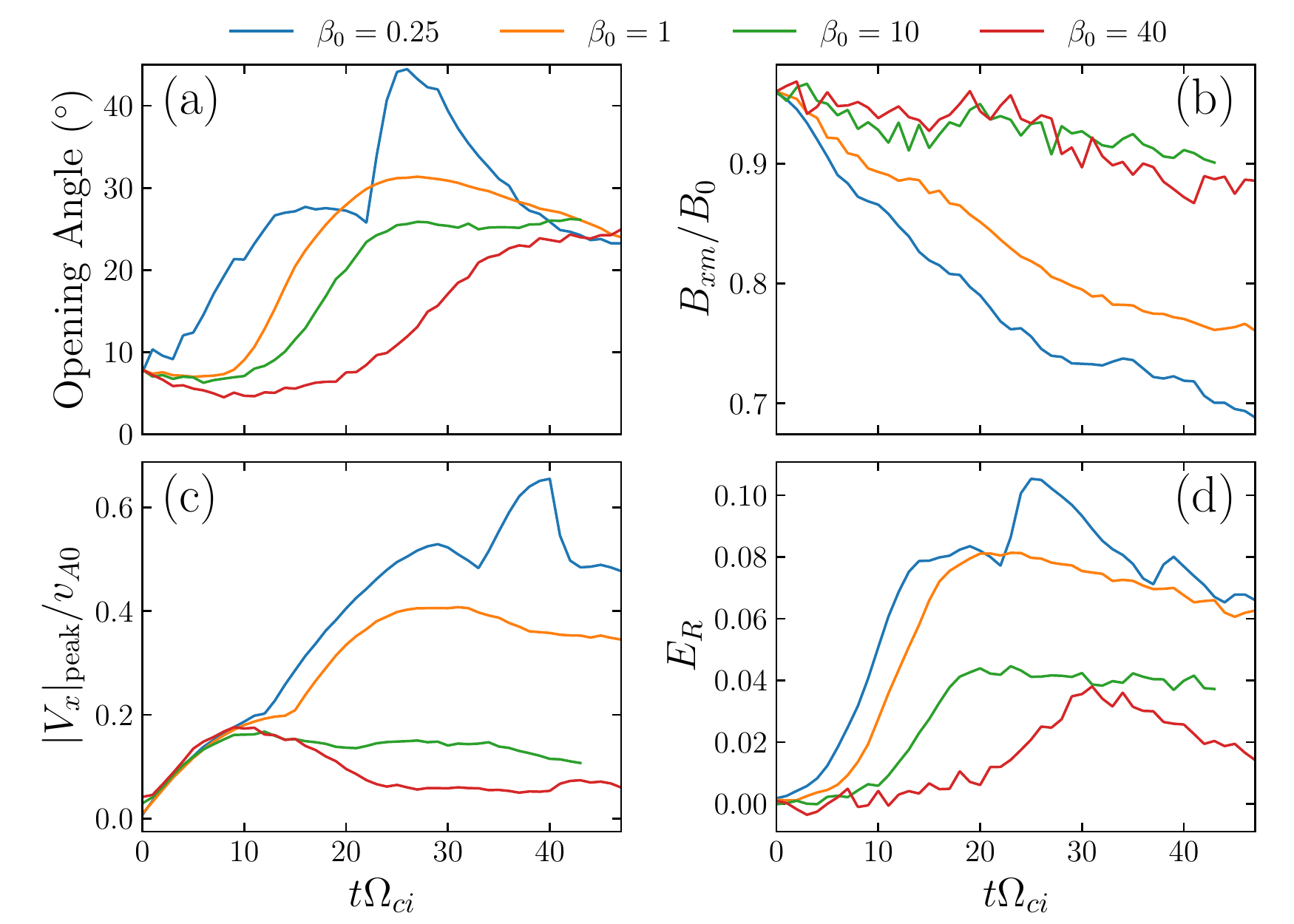}
  \caption{\label{fig:simul}
    Time evolution of key physical quantities in the simulations.
    (a) The opening angle of the reconnection exhaust.
    (b) The magnetic field immediately upstream of the diffusion region. We obtain $B_{xm}$
    using the $B_x$ profile along $z$ and following a similar procedure in
    \citet{Liu2017Why} (see Appendix~\ref{app:bxm} for the detailed procedure).
    (c) Peak ion outflow velocity $V_{ix}$ normalized by the upstream Alfv\'en
    speed $v_{A0}$. We pick the maximum $V_{ix}$ near the X-line in regions between
    the X-point and $10d_i$ downstream of the X-point to estimate the $V_{ix}$ downstream of the IDR. For lower-$\beta$ runs, $V_{ix}$ can still grow further away from the X-point.
    (d) Normalized reconnection rate $E_R$.
  }
\end{figure}

Figure~\ref{fig:simul} shows the time evolution of several quantities critical to the determination of the reconnection rate. Fig.~\ref{fig:simul}(a) shows the exhaust opening angle. The reconnection exhaust opens up earlier in the low-$\beta$ runs, corresponding to a faster reconnection onset in the low-$\beta$ limit. The opening angles in the nonlinear stage fall between 20$^\circ$ and 35$^\circ$ in general. When a secondary plasmoid forms in the reconnection layer (Fig.~\ref{fig:absj_vout}(a)), the opening angle can reach 45$^\circ$ in the $\beta=0.25$ case, but this transient feature between time $20<t\Omega_{ci}<30$ is not our focus. As the exhaust opens up, the upstream-pointing magnetic tension force gets stronger in the inflow region. To maintain this field geometry, the upstream magnetic field strength needs to decrease toward the diffusion region so that the magnetic pressure gradient balances the tension force~\citep{Liu2017Why}. The resulting reconnecting field immediately upstream of the diffusion region $B_{xm}$ will thus be reduced; here the subscript ``m'' denotes the {\it microscopic} scale, which will be the ion inertial scale in electron-proton plasmas. Fig.~\ref{fig:simul}(b) shows the $B_{xm}$ evolution in different runs. This $B_{xm}$ gradually decreases as reconnection proceeds and saturates after $t\Omega_{ci}^{-1}>30$ for all runs. Its value is larger when $\beta$ is higher---about 0.73 when $\beta=0.25$ and 0.9 when $\beta=40$ at $t\Omega_{ci}=30$. Based on this observation, one may expect that the outflow speed $V_{ix}$ in high-$\beta$ runs to be higher since $V_{ix}\propto B_{xm}$, i.e., the outflow is driven by $B_{xm}$. However, as shown in Fig.~\ref{fig:simul}(c), the outflow speed $V_x$ turns out to be lower in high-$\beta$ runs. $V_{ix}$ goes up to $0.5v_{A0}$ when $\beta=0.25$ and decreases below $0.1v_{A0}$ when $\beta=40$, suggesting that other factors besides $B_{xm}$ play important roles in determining the outflow speed. Since the outflow is slower in the high-$\beta$ limit, we expect a lower reconnection rate in this limit. Fig.~\ref{fig:simul}(d) shows that the reconnection rate indeed decreases with plasma $\beta$, as expected. When $\beta\leq1$, the normalized reconnection rate $E_R\equiv cE_y/B_{x0}v_{A0}$ is around 0.08 and can be larger than 0.1 when the secondary plasmoids temporarily widen the reconnection exhaust in the $\beta$=0.25 case, consistent with earlier simulations~\citep[e.g.,][]{Birn2001Geospace}. In contrast, when $\beta=40$, $E_R$ is below 0.04. In the next section, we will develop a model to explain these simulated trends.

\section{Rate model with thermal correction}
\label{sec:rate_model}
\begin{figure}[ht!]
  \centering
  \includegraphics[width=0.6\textwidth]{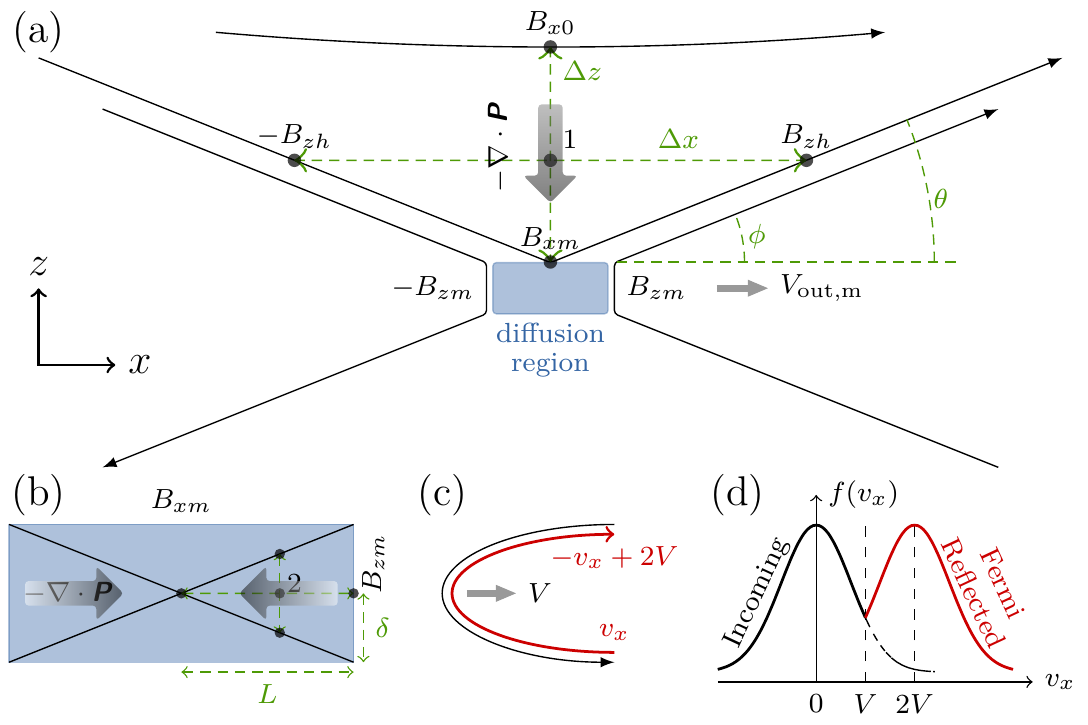}
  \caption{\label{fig:schematic}
    The reconnection rate model with thermal corrections.
    (a) The geometry of reconnection at the mesoscale. The large grey arrow
    indicates the force $-\nabla\cdot\tensorsym{P}$ due to thermal correction.
    The force balance along the inflow will be evaluated at point 1.
    $B_{x0}$ is the asymptotic magnetic field. $B_{xm}$ is the magnetic field
    immediately upstream of the diffusion region. $B_{zh}$ is evaluated near
    the separatrix. $B_{zm}$ is the magnetic field immediately downstream the
    diffusion region. Angles $\theta\equiv\tan^{-1}(\Delta z/\Delta x)$ and
    $\phi\equiv\tan^{-1}(B_{zm}/B_{xm})$.
    (b) Dimensions of the diffusion region $2L\times 2\delta$ at the microscale. The arrows indicate
    the forces toward the X-point arising from the thermal correction.
    The force balance along the outflow will be evaluated at point 2.
    (c) Plasma heating associated with the Fermi mechanism.
    The red curve indicates an example particle trajectory. Real particle trajectory might be more complicated~\citep{Drake2009Ion}. $v_x$ is particle's
    velocity along $x$ when it enters the diffusion region. The flow velocity $V$
    increases from 0 at the X-point to $V_\text{out,m}$ immediately out of the diffusion region.
    (d) Particle velocity distribution consisting of the incoming population and the
    Fermi reflected population.
  }
\end{figure}

To explain the simulation results, we develop the framework to incorporate thermal pressure effects in the reconnection rate model. These effects are expected to be important when $\beta \gtrsim \mathcal{O}(1)$. Figure~\ref{fig:schematic} is an overview of this model, which includes the thermal correction $-\nabla\cdot\tensorsym{P}$ that participates in the force balance. As shown in the schematic, the thermal correction might change the inflow force balance and therefore, $B_{xm}$, and the outflow speed can be slowed down by the back-pressure and/or a weaker magnetic tension due to the pressure anisotropy. Following the approach in~\citet{Liu2017Why}, the force balance evaluated at point 1 in the inflow region will relate $B_{xm}$ to the upstream asymptotic field $B_{x0}$. By matching the upstream magnetic field opening angle $\theta$ to angle $\phi$ made by the reconnected field at the microscopic scale, we obtain the strength of the reconnected field $B_{zm}\simeq B_{xm}\Delta z/\Delta x$.
On the other hand, the force balance evaluated at point 2 within the diffusion region will determine the outflow velocity $V_\text{out,m}$ immediately downstream of the diffusion region. With both $B_{zm}$ and $V_\text{out,m}$, we can calculate the motional electric field adjacent to the diffusion region, which is essentially the reconnection electric field. An expression of the normalized reconnection rate as a function of the opening angle $\theta$ then can be derived to be $E_R \equiv cE_y/B_{x0}v_{A0}=B_{zm}V_\text{out,m}/(B_{x0}v_{A0})$.

Assuming that the pressure tensor can be approximated into the gyrotropic form, then the momentum equation can be written as
\begin{equation}
  nm_i\frac{\partial\vect{V}}{\partial t}+nm_i\vect{V}\cdot\nabla\vect{V} + \nabla\frac{B^2}{8\pi} + \nabla P_\perp
  \simeq\nabla\cdot\left(\varepsilon \frac{\vect{B}\vect{B}}{4\pi}\right),
  \label{equ:force}
\end{equation}
where the anisotropy (firehose) parameter $\varepsilon\equiv1-4\pi(P_\parallel-P_\perp)/B^2$. $P_\|$ and $P_\perp$ are the pressure parallel and perpendicular to the local magnetic field, respectively. The plasma is firehose unstable (e.g., \cite{CPWang20a,CPWang18a}) if $\varepsilon < 0$ since the magnetic tension in the right-hand side becomes negative. When $P_\parallel > P_\perp$, the pressure anisotropy will reduce the magnetic tension force. We seek a steady-state solution by neglecting the $\partial t$ term. From simulations, we get the following empirical relations $P_{e\parallel}\approx P_{e\perp}\approx P_{e0}n/n_0$, $P_{i\parallel} \approx P_{i0}n/n_0$, and $P_{i\perp} \approx P_{i0}nB_x/(n_0B_{x0})$ along the inflow symmetry line (see Appendix~\ref{app:inflow} for details, where the CGL~\citep{CGL1956} and Le \& Egedal~\citep{Le2009,Egedal2013Review} closures are also explored). The simulations further suggest that the plasma number density $n\approx n_0[1-C(1-\bar{B}_x)]$, where the constant $C=1/(\sqrt{2\beta_{i0}}+1)$ (see Fig.~\ref{fig:nbx} for details). By discretizing Eq.(\ref{equ:force}) along the inflow direction at point 1 and using these pressure closures, we get

\begin{equation}
  \frac{B_{x0}^2-B_{xm}^2}{8\pi\Delta z}
  + \frac{P_{0}+P_{i0}(\sqrt{2\beta_{i0}}+B_{xm}/B_{x0})}{\sqrt{2\beta_{i0}}+1}\frac{B_{x0}-B_{xm}}{B_{x0}\Delta z}
  = \frac{\varepsilon_1}{4\pi}B_{x1}\frac{2B_{zh}}{\Delta x}.
  \label{equ:bxm_o}
\end{equation}
 We made a reasonable approximation of $B_x\simeq B_x(z)$, $B_z\simeq B_z(x)$, and $\varepsilon\simeq\varepsilon(z)$ around the diffusion region. The first term is from the magnetic pressure gradient, the second term is from the perpendicular pressure gradient, and the term on the right is from the magnetic tension force modified by the pressure anisotropy. $B_{x1}=(B_{x0}+B_{xm})/2$ is the magnetic field at point 1 and the corresponding firehose parameter is
 \begin{align}
  \varepsilon_1 \equiv \varepsilon(B_{x1}) = 1 + \frac{\beta_{i0}}{2}\left(\frac{1}{\bar{B}_{x1}} - \frac{1}{\bar{B}_{x1}^2}\right)\left(1-\frac{1-\bar{B}_{x1}}{\sqrt{2\beta_{i0}}+1}\right),
  \label{equ:epsilon1}
\end{align}
where $\bar{B}_{x1}\equiv B_{x1}/B_{x0}$. Note that we can include the plasma inertia ($nm_i\vect{V}\cdot\nabla\vect{V}$) in the calculation, but the inflow contribution is negligible (i.e., can be ordered out), thus we will not treat it here to avoid unnecessary complexity. Equation~(\ref{equ:bxm_o}) can be written in a dimensionless form (see Appendix~\ref{app:inflow} for details)
\begin{align}
  (1 - \bar{B}_{xm}^2) + \frac{\beta_{0}+\beta_{i0}(\sqrt{2\beta_{i0}}+\bar{B}_{xm})}{\sqrt{2\beta_{i0}}+1}\left(1-\bar{B}_{xm}\right)
  = \varepsilon_1\left(\frac{\Delta z}{\Delta x}\right)^2
  \left(1+\bar{B}_{xm}\right)^2,
  \label{equ:Bxm}
\end{align}
where $\bar{B}_{xm}\equiv B_{xm}/B_{x0}$ and the asymptotic upstream plasma $\beta_0\equiv \beta_{i0}+\beta_{e0}$. In cases studied here $\beta_{i0}=\beta_{e0}=8\pi n_0kT_0/B_{x0}^2$. When the opening angle is small ($\Delta z/\Delta x\ll1$), $\bar{B}_{xm}\to 1$ for any $\beta_0$, consistent with that early in the simulations (Fig.~\ref{fig:simul}(b)). For an arbitrary opening angle, we can obtain $\bar{B}_{xm}$ numerically finding the roots of Eq.~(\ref{equ:Bxm}) (e.g., using Newton's method).

To obtain the outflow velocity, we follow a similar procedure as in the inflow region but discretize the force-balance along the x-direction at point 2 in Fig.~\ref{fig:schematic}(b) and keep the plasma inertia ($nm_i\vect{V}\cdot\nabla\vect{V}$).  We get
\begin{align}
  \frac{n_2m_i V_\text{out,m}^2}{2L} + \frac{B_{zm}^2}{8\pi L} +
  \frac{\Delta P_{xx,m}}{L}
  & = \frac{1}{4\pi}\frac{B_{zm}}{2}\frac{\varepsilon_mB_{xm}/2}{\delta/2},
  \label{equ:balance_outflow}
\end{align}
where $n_2$ is the plasma density at point 2, $\Delta P_{xx,m}$ is the pressure increases from the X-point to immediately downstream of the ion diffusion region, and $\varepsilon_m$ is the anisotropy parameter (similar to Eq.~(\ref{equ:epsilon1})) immediately upstream of the ion diffusion region
\begin{equation}
  \varepsilon_m\equiv \varepsilon(B_{xm})
  = 1 + \frac{\beta_{i0}}{2}\left(\frac{1}{\bar{B}_{xm}} - \frac{1}{\bar{B}_{xm}^2}\right)\left(1-\frac{1-\bar{B}_{xm}}{\sqrt{2\beta_{i0}}+1}\right).
  \label{equ:epsilon_m}
\end{equation}
which depends on $\beta_{i0}$ and $\bar{B}_{xm}$ and can be determined once we obtain $\bar{B}_{xm}$ from Eq.~(\ref{equ:Bxm}). To find the solutions of $V_\text{out,m}$ from Eq.~(\ref{equ:balance_outflow}), we need to model $n_2$ and $\Delta P_{xx,m}$. Unlike the inflow region, the CGL-like closure is not expected to work within the ion diffusion region. To estimate $\Delta P_{xx,m}$, we only need to know the difference between $P_{xx}$ at the x-line and the edge of the ion diffusion region. While the particle heating mechanism can be complex inside the diffusion region \citep{Hoshino2001Suprathermal,Shuster2015,Wang2016Electron}, it transitions to a simpler Fermi-type reflection outside the diffusion region. We thus will model the $P_{xx,m}$ using Fermi-mechanism, as illustrated in Figs.~\ref{fig:schematic}(c) \& (d); particle velocity changes from $v_x$ to $2V-v_x$ during the reflection (Fig.~\ref{fig:schematic}(c)), and the combination of the reflected/accelerated particles with the incoming population leads to the broadening of the distribution function (Fig.~\ref{fig:schematic}(d), see Appendix~\ref{app:outflow} for details). Strictly speaking, the heating is primarily along the magnetic field, but particles can be scattered near the diffusion region, leading to an increase of the perpendicular pressure $\Delta P_\perp$, which for ions is approximately 
\begin{align}
  \Delta P_{ixx}(V) = n_0m_i\left[V^2
  + \left(V^2+\frac{\beta_{i0}}{2}v_{A0}^2\right)
  \erf\left(\frac{V}{\sqrt{\beta_{i0}}v_{A0}}\right)
  + Vv_{A0}\sqrt{\frac{\beta_{i0}}{\pi}}e^{-V^2/(\beta_{i0}v_{A0}^2)}\right].
  \label{equ:Fermi}
\end{align}
The corresponding plasma density is
\begin{align}
  n(V) = n_0 + n_0\erf\left(\frac{V}{\sqrt{\beta_{i0}}v_{A0}}\right),
  \label{equ:density}
\end{align}
which approaches $n_0$ when $\beta_{i0}\to\infty$ and $2n_0$ when $\beta_{i0}\to 0$, within the range predicted in \cite{Birn2010Scaling}. The contribution from electrons through a single Fermi-reflection is negligible because $\Delta P_{exx}/\Delta P_{ixx}\sim\mathcal{O}(\sqrt{m_e/m_i}) \ll 1$ according to Fig.(\ref{equ:Fermi}). Even though the ion $\Delta P_{ixx}$ may not totally account for $\Delta P_{xx}$, we find it remains the dominant term even in the large $\beta$ limit (shown in Sec. 4). Observationally, it has been shown that ion heating is stronger than electron heating during magnetic reconnection at Earth's magnetosphere, and $\Delta T_i/\Delta T_e\simeq 7$~\citep{Phan2013Electron,Phan2014Ion}. Encouraged by these observations, we will ignore electron heating $\Delta P_{exx}$ in the following analysis. Other potential heating mechanisms~\citep{Hoshino2001Suprathermal,Drake2005Production,Drake2006Electron,Oka2010Electron,Egedal2015Double,Shuster2015,Dahlin2014Mechanisms,Li2015Nonthermally,Wang2016Electron} could be included in future work. In the following analysis, we will take $\Delta P_{xx,m}\simeq\Delta P_{ixx,m}\equiv\Delta P_{ixx}(V_\text{out,m})$, which is normalized to
\begin{align}
  \Delta\bar{P}_{ixx,m} = \frac{4\pi\Delta P_{ixx,m}}{B_{x0}^2}
  = \bar{V}_\text{out,m}^2 +
  \left(\bar{V}_\text{out,m}^2 + \frac{\beta_{i0}}{2}\right)
  \erf\left(\frac{\bar{V}_\text{out,m}}{\sqrt{\beta_{i0}}}\right) +
  \bar{V}_\text{out,m}\sqrt{\frac{\beta_{i0}}{\pi}}e^{-\bar{V}_\text{out,m}^2/\beta_{i0}},
  \label{equ:dpixx_m}
\end{align}
where $\bar{V}_\text{out,m}=V_\text{out,m}/v_{A0}$ and we have used $v_{A0}=B_{x0}/(4\pi n_0m_i)^{1/2}$. We model $n_2$ as the average of the density at the X-point ($\simeq n_0$) and that immediately downstream of the ion diffusion region ($\simeq n(V_\text{out,m})$ evaluated from Eq.~(\ref{equ:density})). Thus,
\begin{align}
  n_2 \simeq \frac{n_0 + n(V_\text{out,m})}{2} = n_0 + \frac{n_0}{2}\erf\left(\frac{\bar{V}_\text{out,m}}{\sqrt{\beta_{i0}}}\right).
\end{align}
Dividing Eq.~(\ref{equ:balance_outflow}) by $n_0m_iv_{A0}^2$, we get the force balance at point 2 (see Appendix~\ref{app:outflow} for details)
\begin{align}
  \frac{1}{4}\left[2+\erf\left({\frac{\bar{V}_\text{out,m}}{\sqrt{\beta_{i0}}}}\right)\right]
  \bar{V}_\text{out,m}^2 + \Delta\bar{P}_{ixx,m} +
  \frac{\bar{B}_{xm}^2}{2}\left[\left(\frac{\Delta z}{\Delta x}\right)^2 -
  \varepsilon_m\right] \simeq 0
  \label{equ:Voutm}
\end{align}
in dimensionless form, where we have used $B_{zm}/B_{xm}\simeq\delta/L\simeq\Delta z/\Delta x$.

When $\beta_0\ll 1$, then $\erf(\bar{V}_\text{out,m}/\sqrt{\beta_{i0}})\to 1$ and $\bar{V}_\text{out,m}\exp(-\bar{V}_\text{out,m}^2/\beta_{i0})\to 0$. As a result, $\Delta\bar{P}_{ixx,m}\to 2\bar{V}_\text{out,m}^2$ according to Eq.~(\ref{equ:dpixx_m}). In the small opening angle limit, $\Delta z/\Delta x\ll 1$, $\bar{B}_{xm}\to 1$ according to Eq.~(\ref{equ:Bxm}), and $\varepsilon_m\to 1$ according to Eq.~(\ref{equ:epsilon_m}). Then,
\begin{equation}
  V_\text{out,m} = \sqrt{\frac{2\varepsilon_m}{11}}v_{A0}\simeq 0.43v_{A0},
\end{equation}
which is an Alfv\'enic outlfow as expected in the low-$\beta$ limit.

When $\beta_0\gg 1$, then $\erf\left(\bar{V}_\text{out,m}/\sqrt{\beta_{i0}}\right)\simeq (2/\sqrt{\pi})\bar{V}_\text{out,m}/\sqrt{\beta_{i0}}$ and $\bar{V}_\text{out,m}\exp(-\bar{V}_\text{out,m}^2/\beta_{i0}) \simeq \bar{V}_\text{out,m}$. According to Eq.~(\ref{equ:dpixx_m}),
\begin{equation}
  \Delta\bar{P}_{ixx,m}\to\frac{2}{\sqrt{\pi\beta_{i0}}}\bar{V}_\text{out,m}^3 + \bar{V}_\text{out,m}^2 + 2\sqrt{\frac{\beta_{i0}}{\pi}}\bar{V}_\text{out,m}.
\end{equation}
Then, Eq.~(\ref{equ:Voutm}) gives
\begin{equation}
  \frac{5}{\sqrt{\pi\beta_{i0}}}\bar{V}_\text{out,m}^3 + 3\bar{V}_\text{out,m}^2 + 4\sqrt{\frac{\beta_{i0}}{\pi}}\bar{V}_\text{out,m} = \varepsilon_m,
  \label{vout_eqn_hbeta}
\end{equation}
where $0\leq\varepsilon_m\leq 1$. We seek a solution $\bar{V}_\text{out,m}>0$ (flows moving away from the X-point). All the terms on the left-hand side of Eq.~(\ref{vout_eqn_hbeta}) are positive, indicating that the third term $4\sqrt{\beta_{i0}/\pi}\bar{V}_\text{out,m}<\varepsilon_m$. Then, the first term $5\bar{V}_\text{out,m}^3/\sqrt{\pi\beta_{i0}}<(5\pi/(64\beta_{i0}^2))\varepsilon_m^3$ and the second term $3\bar{V}_\text{out,m}^2 < (3\pi/(16\beta_{i0}))\varepsilon_m^2$. Both terms will be much smaller than $\varepsilon_m$ when $\beta_{i0}\gg 1$ 
and are, therefore, small corrections to the first-order linear equation $4\sqrt{\beta_{i0}/\pi}\bar{V}_\text{out,m} = \varepsilon_m$. Thus,
\begin{equation}
  V_\text{out,m}\simeq \frac{\sqrt{\pi}}{4}\frac{\varepsilon_mv_{A0}}{\sqrt{\beta_{i0}}}.
  \label{vout_hbeta}
\end{equation}
A higher $\beta$ will thus reduce the outflow speed. Equation (\ref{vout_hbeta}) is almost identical to the expression obtained in \citet{Haggerty2018}, that is $(\sqrt{2}/3)\varepsilon_mv_{A0}/\sqrt{\beta_{i0}}$ in our notation. They compared this expression against with 81 kinetic simulations and 14 in-situ observations that span a wide range of parameter regimes, and showed an excellent agreement. However, the $\sqrt{2}/3$ factor is an empirical parameter in their model while we derived it from Eq.~(\ref{vout_eqn_hbeta}). It is also interesting to note that their prediction based on the 1D shock transition across the exhaust is consistent with our 2D model that accounts for the back-pressure along the outflow direction. The single compact expression in Eq.(\ref{equ:Voutm}) explains the outflow speed in both the high-$\beta$ and low-$\beta$ limit reported in \citet{Haggerty2018}. For an arbitrary plasma $\beta$ and opening angle, we can find the roots of Eq.(\ref{equ:Voutm}) numerically. Combining Eqs.(\ref{equ:Bxm}), (\ref{equ:Fermi}), (\ref{equ:Voutm}) and $B_{zm}\simeq B_{xm}\Delta z/\Delta x$, the resulting reconnection rate $E_R= B_{zm}V_\text{out,m}/(B_{x0}v_{A0})$ can be derived for a general case.

\section{Model-simulation comparison}
\label{sec:compare}

\begin{figure}[ht!]
  \centering
  \includegraphics[width=0.5\textwidth]{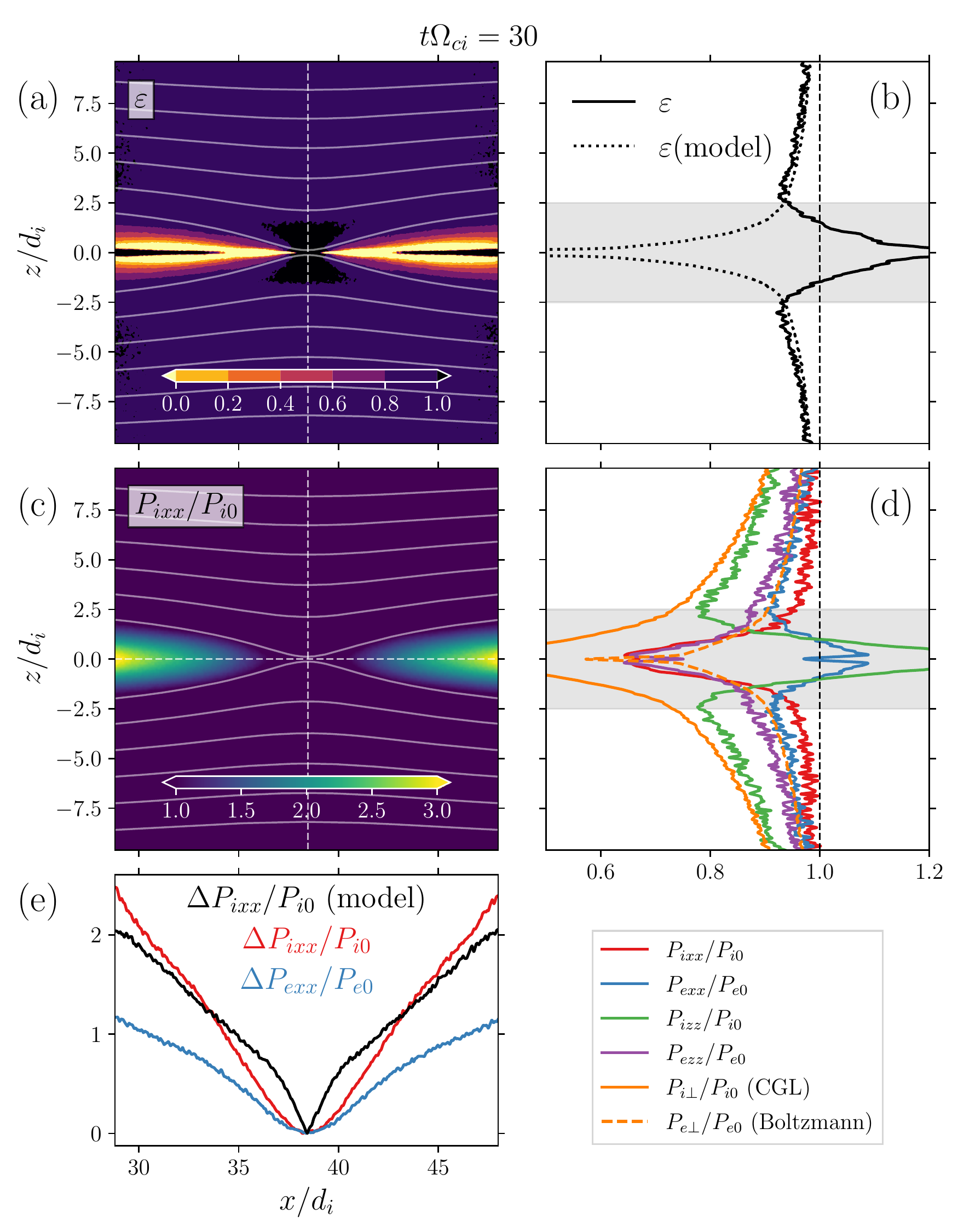}
  \caption{\label{fig:thermal}
    Thermal effects in the run with $\beta=1$.
    (a) The firehose parameter $\varepsilon$.
    (b) $\varepsilon$ along the vertical cut across the X-point (the dashed line in (a)). The dotted line is the modeled $\varepsilon$. The shaded region indicates the ion diffusion region.
    (c) $P_{ixx}$ normalized by the background pressure.
    (d) Vertical cuts of the electron and ion pressure tensor components. The orange curves show the CGL (solid) and Boltzmann (dashed) scalings. The shaded region indicates the ion diffusion region.
    (e) Ion pressure enhancement along $z=0$ (along the horizontal dashed line in (c)). The black line shows the predicted $\Delta P_{ixx}$ heating from the Fermi mechanism, which is evaluated from Eq.~(\ref{equ:Fermi}) using measured $V_{ix}$ along $z=0$. For reference, the blue curve shows the enhancement of electron pressure $\Delta P_{exx}$.
  }
\end{figure}

To illustrate the thermal effects in simulations, we plot the anisotropy (firehose) parameter $\varepsilon$ and relevant components of the pressure tensors for the $\beta=1$ case in Fig.~\ref{fig:thermal}.
We first examine quantities important to the force-balance upstream of the ion diffusion region (IDR). The 2D map of $\varepsilon$ near the IDR is shown in Fig.~\ref{fig:thermal}(a), and its (vertical) cut across the x-line along the inflow direction is shown in Fig.~\ref{fig:thermal}(b). The model based on the closure we choose (dotted black) captures the decreasing trend of $\varepsilon$ toward the IDR, which is shaded in grey. Note that the region inside the grey area is not critical to the upstream force-balance discussed here.
This anisotropy parameter $\varepsilon$ (solid black) is calculated using $P_\|$ and $P_\perp$ in the simulation, and along the inflow symmetry line $P_\|\simeq P_{xx}$ and $P_\perp\simeq P_{zz}$. Therefore, in Fig.~\ref{fig:thermal}(d) we show the vertical cuts of $P_{ixx}$, $P_{izz}$, $P_{exx}$ and $P_{ezz}$. The perpendicular component $P_{izz}$ follows well with the CGL closure based on the $\mu$-conservation (solid orange). $P_{ixx}$ only decreases slightly and is closer to the Boltzmann closure (dashed orange). Both $P_{exx}$ and $P_{ezz}$ also follow the Boltzmann closure better, resulting in a much smaller pressure anisotropy than ions.
In addition to the magnetic tension reduction due to $\varepsilon$, the pressure gradient $\partial_z P_{zz}$ resulting from the $P_{zz}$ drop also plays an important role in counter-balancing the upstream-pointing tension force, mitigating the decrease of $B_{xm}$.

Along the outflow, the back-pressure from $P_{xx}$ gradient is especially critical in the force-balance. We thus plot the 2D map of the dominant component $P_{ixx}$ in Fig.~\ref{fig:thermal}(c) and its (horizontal) cut across the x-line in Fig.~\ref{fig:thermal}(e). Our model $\Delta P_{ixx}$ (Eq.(\ref{equ:Fermi}) in black) based on the Fermi reflection reasonably captures the increasing trend toward the downstream region. In addition to this back-pressure, $\varepsilon$ upstream of the IDR (shown in panel (b)) can further reduce the magnetic tension force that drives the reconnection outflow. Note that this is consistent with the Wal\'en test \citep{Sonnerup1981} across exhausts, which indicates the upstream $\varepsilon$ (instead of the downstream $\varepsilon$) can affect the outflow velocity. This fact is also captured in the $\varepsilon_m$ dependence in our model Eq.(\ref{equ:Voutm}). Fig.~\ref{fig:thermal}(e) also shows a significant $\Delta P_{exx}$, likely resulting from other heating mechanisms~\citep{Hoshino2001Suprathermal,Drake2005Production,Drake2006Electron,Oka2010Electron,Egedal2015Double,Shuster2015,Dahlin2014Mechanisms,Li2015Nonthermally,Wang2016Electron} not considered here. However, simulations suggest that ion heating remains dominant (i.e., $\Delta P_{ixx} \gtrsim \Delta P_{exx}$) even in the high-$\beta$ limit, and the goal of this paper is to lay out a framework capable of incorporating the thermal correction on reconnection rates. A future more sophisticated and accurate heating model could be included in a similar manner.

\begin{figure}[ht!]
  \centering
  \includegraphics[width=0.6\textwidth]{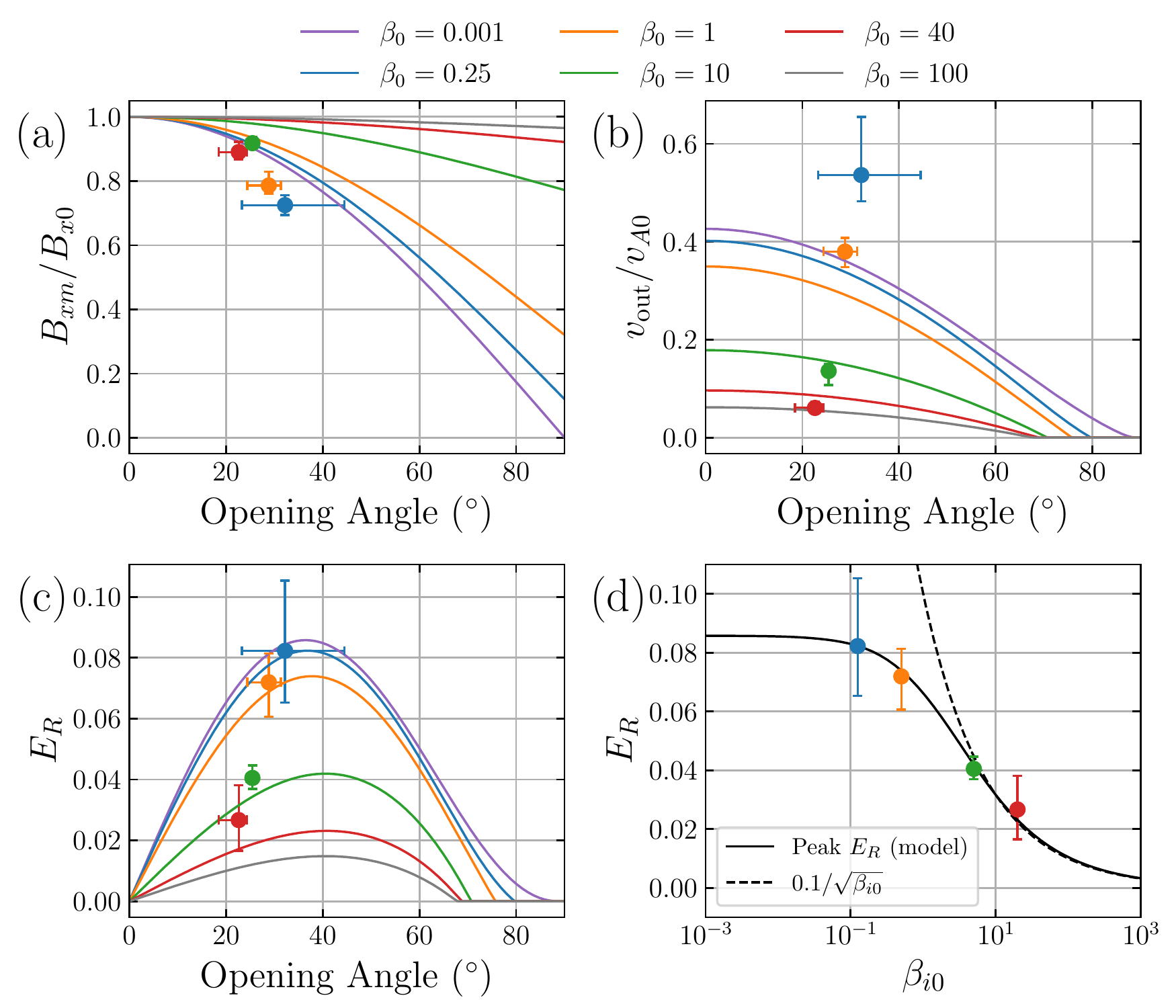}
  \caption{\label{fig:model}
    Model predictions using Eqs.~(\ref{equ:Bxm}), (\ref{equ:Fermi}), (\ref{equ:Voutm}) and $E_R= B_{zm}V_\text{out,m}/(B_{x0}v_{A0})$. Here $\beta_{i0}=\beta_0/2$. The symbols indicate the simulation data points. We average the simulation results in Fig.~\ref{fig:simul} after the reconnection rate peaks and use the minimum and maximum values in the same time range to determine the error bars.
    (a) The magnetic field upstream the diffusion region.
    (b) Outflow velocity. (c) Normalized reconnection rate.
    (d) The scaling of the predicted maximum reconnection rate with $\beta_{i0}$.
  }
\end{figure}

These pressure models are included in the force-balance along the inflow (Eq.~(\ref{equ:Bxm})) to obtain the strength of reconnecting field ($B_{xm}$) immediately upstream of the diffusion region and along the outflow (Eq.(\ref{equ:Voutm})) to obtain the velocity ($V_\text{out,m}$) leaving the ion diffusion region. Fig.~\ref{fig:model}(a) shows the modeled $B_{xm}$ as a function of the opening angle $\theta$ for cases with a wide range of upstream plasma $\beta_0=10^{-3}-10^2$. In the small opening angle limit, $B_{xm}\rightarrow B_{x0}$, as in an elongated Sweet-Parker reconnection layer. With the large opening angle, magnetic pressure $\partial_z B^2/8\pi$ is required to balance the upstream-pointing tension force, and this reduces $B_{xm}$; this reduction can be mitigated in the large-$\beta$ limit since (1) the thermal pressure gradient $\partial_z P_{zz}$ helps balance the magnetic tension and (2) the temperature anisotropy $\varepsilon$ reduces the magnetic tension. This trend is observed in the simulation data ($\beta_0=0.25,1,10,40$) overlaid in the same plot, although this model has overestimated $B_{xm}$. 

Fig.~\ref{fig:model}(b) shows the modeled $V_\text{out,m}$ as a function of the opening angle $\theta$. With a large opening angle, the outflow speed can be reduced because of a weaker $B_{xm}$ and larger (magnetic) back-pressure $\partial_x B^2/8\pi$. The outflow velocity can be further reduced in the large-$\beta$ limit because of (1) the thermal back-pressure $\partial_x P_{xx}$ and (2) the reduced magnetic tension by the temperature anisotropy $\varepsilon$. There are critical angles above which the outflow speed vanishes; this cut-off behavior is caused by the complete loss of magnetic tension (i.e., the driver of reconnection) when $\varepsilon_m < 0$ in the large opening angle limit. This trend of outflow reduction is also observed in the simulation data with different $\beta$, although this model has underestimated $V_\text{out,m}$ in the low-$\beta$ limit.

The resulting reconnection rate is then $E_R=B_{zm}V_\text{out,m}/(B_{x0}v_{A0})$ and the reconnected field strength $B_{zm}\simeq B_{xm}\tan\theta$. Fig.~\ref{fig:model}(c) shows the modeled rate as a function of the opening angle $\theta$. The model predicts a lower reconnection rate with a higher plasma $\beta$; this trend is again observed in the simulation data. Note that the predicted peak $E_R$ state has $\theta$ within $[30^\circ, 40^\circ]$ for a wide range of $\beta$, which is consistent with the $\beta$-insensitive opening angle observed in simulations. Fig.~\ref{fig:model}(d) plots the predicted peak $E_R$ as a function of upstream $\beta_{i0}$, and it captures the decreasing trend of simulated reconnection rate with a larger $\beta_{i0}$. In the $\beta_{i0} \gg \mathcal{O}(1)$ limit, the reduction of reconnection rate primarily correlates with the outflow speed reduction since $B_{xm}$ is not reduced much in this limit (panel (a)). This leads to $\varepsilon_m\simeq 1$, and the outflow speed is thus $V_\text{out,m}\simeq (\sqrt{\pi}/4)(v_{A0}/\sqrt{\beta_{i0}})$ from Eq.~(\ref{vout_hbeta}).
Overall, the predicted reconnection rates show good agreement with simulations.
In this high-$\beta$ limit, the predicted scaling of the maximum plausible reconnection rate can be well approximated as $E_R\simeq 0.1/\sqrt{\beta_{i0}}$, as indicated by the dashed black curve in Fig.~\ref{fig:model}(d).

\section{Conclusion and Discussion}
\label{sec:con}
In this paper, we derive an analytical framework to incorporate thermal effects in the reconnection rate model~\citep{Liu2017Why}. These thermal effects are manifested as the pressure anisotropy and pressure gradient force, which can modify the force-balance at both the inflow and outflow regions. In the large-$\beta$ limit, we find that the reconnection rate decreases primarily because of the reduction of the outflow speed, hindered by the pressure gradient force. The spatial variation of thermal pressure is modeled using a combination of CGL and Boltzmann closures in the upstream and the kinetic heating by Fermi reflections in the downstream. The pressure gradient force derived from the Fermi-heating only depends on the upstream plasma parameters and the outflow speed, and it can be easily included in the rate model. 2D kinetic simulations compare favorably with the heating mechanism and the $\beta$-dependency of various key quantities needed to model the outflow speed and reconnection rate.

While the force-balance constraint laid out here is general in determining the reconnection rate, there are opportunities for improvement.
First, the present model does not include a guide field, which could change the plasma heating~\citep{Dahlin2014Mechanisms,Li2017Particle} and scattering processes, and therefore, the pressure anisotropy and pressure gradient force. As shown by~\citet{Haggerty2018}, the outflow velocity in the reconnection exhaust tends to get closer to $v_{A0}$ as the guide field increases. This is also expected from our model; a background guide field could inhibit the upstream pressure variation due to the $\mu$-conservation since guide-field strength does not change much while convects into the diffusion region; it may also reduce the acceleration rate of Fermi mechanism for $P_{xx}$ heating since the field-line curvature ($\vect{\kappa}$ and thus the heating rate $e_\|(\vect{v}_E\cdot \vect{\kappa}$)) is reduced \citep{Dahlin2014Mechanisms,Li2017Particle}. 
Second, when the ion gyro-radius is much larger than the current layer thickness, the gyrotropic approximation is expected to break, and the full pressure tensor needs to be considered. 
Third, the model only includes ion heating in the diffusion region, while additional electron heating can be as strong in some regimes~\citep{Haggerty2015Competition, Shuster2015,Wang2016Electron,Dahlin2014Mechanisms,Wang2016Electron,Li2015Nonthermally}, and the upstream electric potential may be important~\citep{Le2009,Egedal2013Review,Egedal2015Double,Shuster2015}. 
Fourth, pressure anisotropy instabilities (e.g., mirror instability) may arise in high-$\beta$ plasmas and was suggested to distort the field geometry and the current sheet~\citep{Alt2019Onset}. 
Finally, we do not take 3D physics into account, for example, self-generated turbulence~\citep{Daughton2011Role,Liu2013Bifurcated,Li2019Particle} or localized reconnection layer along the third dimension~\citep{KHuang20a, yhliu19a}. 
Nevertheless, this present model extends the reconnection rate model to the high-$\beta$ regime and provides new insights into the reconnection rate problem in high-$\beta$ plasmas, which can be applicable to the outer heliosphere, hot intracluster medium of galaxy clusters, and the Galactic center.

\acknowledgments

We thank the anonymous referee for very helpful and constructive reviews. We acknowledge support by the National Science Foundation grant PHY-1902867
through the NSF/DOE Partnership in Basic Plasma Science and Engineering and
NASA MMS 80NSSC18K0289. Simulations were performed at National Energy Research
Scientific Computing Center (NERSC), at the Texas Advanced Computing Center
(TACC) at The University of Texas at Austin, and with Los Alamos National Laboratory
(LANL) institutional computing.

\appendix
\section{Determining the reconnecting magnetic field upstream of the diffusion region}
\label{app:bxm}
\begin{figure}[ht!]
  \centering
  \includegraphics[width=0.5\textwidth]{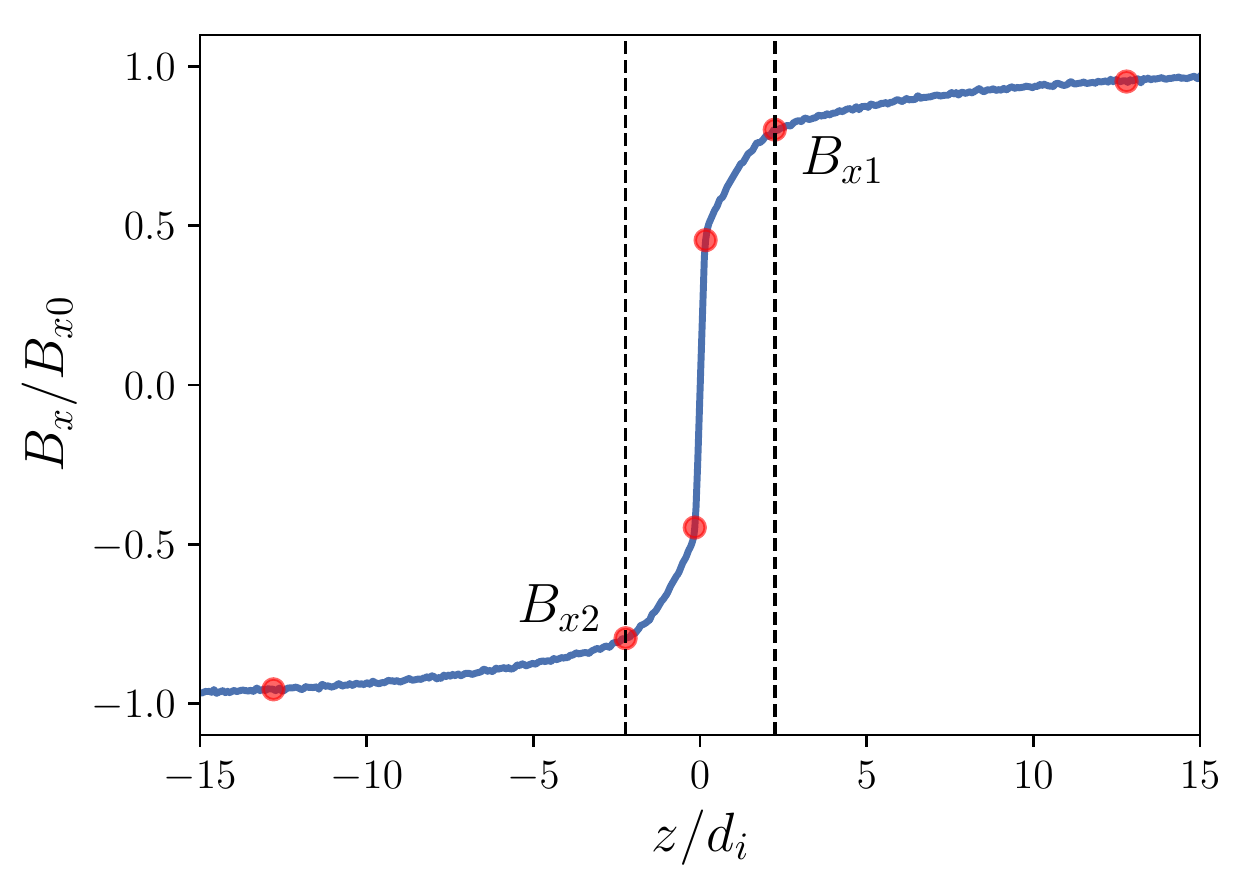}
  \caption{\label{fig:bx_fit}
  $B_x$ profile through the X-line at $t\Omega_{ci}=30$ in the run with $\beta=1$. The red dots are the breakpoints automatically determined from piecewise linear fitting of the $B_x$ profile using the \texttt{pwlf} Python library~\citep{pwlf}. The dashed lines indicate the boundaries of the ion diffusion region (IDR). $B_{x1}$ and $B_{x2}$ indicate the magnetic fields immediately upstream of the IDR. Note that we have applied a Gaussian filter to the $B_x$ profile to reduce the grid-scale noises.
  }
\end{figure}

To determine the reconnecting magnetic field upstream of the ion diffusion region ($B_{xm}$), we need to locate the diffusion region first. In the low-$\beta$ regime, one could locate the ion diffusion region by checking where the total electric field $\vect{E}$ starts to deviate from the ideal electric field $-\vect{V}\times\vect{B}$ in the reconnection inflow region. However, this method does not work well in the high-$\beta$ simulations, where both $\vect{E}$ and $\vect{V}$ are very noisy. In this study, we use only the $B_x$ $z-$profile through the X-line to determine $B_{xm}$. Figure~\ref{fig:bx_fit} shows the procedure. We fit the $B_x$ profile with 5 piecewise linear segments and determine a pair of the breakpoints as the boundaries of the ion diffusion region. Then, we calculate $B_{xm}=(B_{x1}-B_{x2})/2$, where $B_{x2} < 0$.

\section{Inflow force balance}
\label{app:inflow}
\begin{figure}[ht!]
  \centering
  \includegraphics[width=0.5\textwidth]{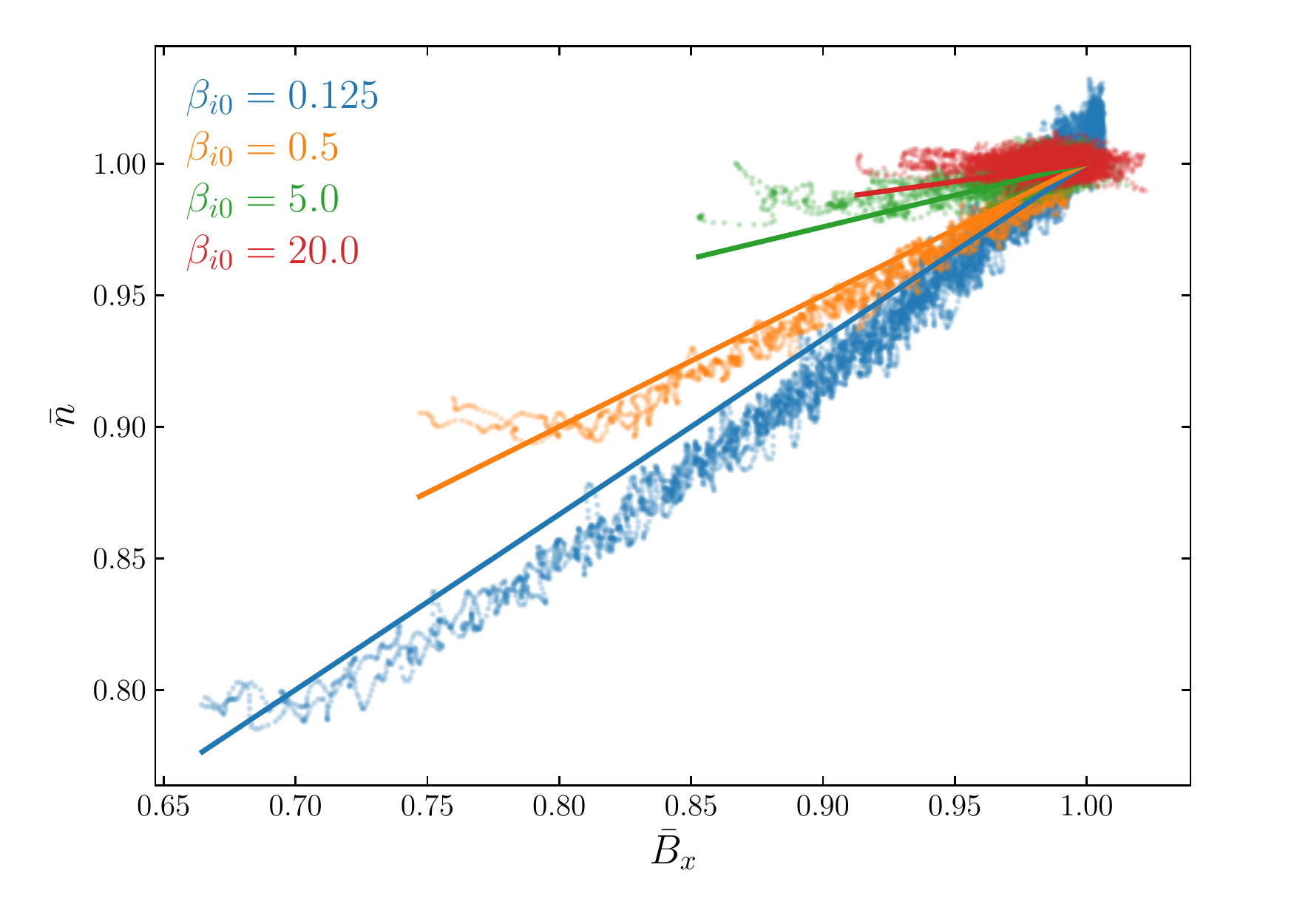}
  \caption{\label{fig:nbx} Plasma density versus the magnetic field strength along the inflow symmetry line for the four runs at $t\Omega_{ci}=30$. $\bar{B}_x=B_x/B_{0}$, where $B_{0}=B_{x0}$ is the asymptotic magnetic field. $\bar{n}=n/n_0$. The straight lines show the linear relation $\bar{n}=1-C(1-\bar{B}_x)$, where the slope $C=1/(\sqrt{2\beta_{i0}}+1)$.
  }
\end{figure}

Starting from the single-fluid momentum equation
\begin{equation}
  \label{equ:mom}
  nm_i\frac{d\vect{V}}{dt} = -\nabla\cdot\tensorsym{P} + \rho\vect{E} +
  \frac{\vect{j}\times\vect{B}}{c},
\end{equation}
we seek for a steady-state solution by neglecting the time-derivative of the plasma inertia. The electric force from charge separation is negligible in the non-relativistic limit. In a well-magnetized plasma, $\tensorsym{P}=P_\perp\tensorsym{I}+
(P_\parallel-P_\perp)\hat{\vect{b}}\hat{\vect{b}}$, where where $P_\parallel$ and
$P_\perp$ are pressures parallel and perpendicular to the local
magnetic field, $\tensorsym{I}$ is the unit dyadic, $\hat{\vect{b}}$ is the unit
vector along the local magnetic field. Then,
\begin{equation}
  \nabla\left(P_\perp + \frac{B^2}{8\pi}\right) = \nabla\cdot\left(\varepsilon
    \frac{\vect{B}\vect{B}}{4\pi}\right),
\end{equation}
where $\varepsilon=1-4\pi(P_\parallel-P_\perp)/B^2$ is the anisotropy (firehose) parameter. $P_\parallel$ and $P_\perp$ will be determined by the local plasma density and magnetic field strength through certain fluid closure. Note that the flux tube tends to expand while it is convected toward the ion diffusion region (i.e., as the result of exhaust opening), leading to a lower plasma density. However, particles can be redistributed along the flux tubes, thus plasma density can change in a slower rate along the inflow symmetry line than the magnetic field does. Fig.~\ref{fig:nbx} shows the relation between the density change and the $B_x$ change along the inflow in our simulations. When $\beta$ is low, and plasma is cold, the redistribution along the field line is less effective, resulting in $\bar{n}\sim\bar{B}_x$. When $\beta$ is high the thermal conduction can be large, thus the redistribution is more effective, resulting in $\bar{n}\to 1$. Simple fitting shows a linear relation $\bar{n}\approx 1-C(1-\bar{B}_x)$ with a constant slope $C=1/(\sqrt{2\beta_{i0}}+1)$. Since $\beta_{i0}$ is varied by changing the thermal temperature in our simulations, this expression indicates that the much heavier ions control the density variation.

The simulation results (e.g., Fig.~\ref{fig:thermal}) show that the electron pressure anisotropy is much smaller than the ion pressure anisotropy. By comparing the results with Boltzmann ($P_\parallel=P_\perp\sim n$), CGL~\citep{CGL1956}, and Le \& Egedal~\citep{Le2009,Egedal2013Review} closures, we find that the pressure terms along the inflow symmetry line are best modeled as
\begin{align}
  & P_{e\parallel}\approx P_{e\perp}=P_{e0}\bar{n}, \\
  & P_{i\parallel}=P_{i0}\bar{n},\quad P_{i\perp}= P_{i0}\bar{n}\bar{B}_x.
\end{align}
Then, the total parallel and perpendicular pressures are  
\begin{align}
  P_\parallel & = P_0\left(1-\frac{1-\bar{B}_x}{\sqrt{2\beta_{i0}}+1}\right)\\
  P_\perp & =(P_{e0}+P_{i0}\bar{B}_x)\left(1-\frac{1-\bar{B}_x}{\sqrt{2\beta_{i0}}+1}\right),
\end{align}
where $P_0=P_{e0}+P_{i0}$ is the total pressure of the background plasma.
By approximating $B_x\simeq B_x(z)$, $B_z \simeq B_z(x)$, and $\varepsilon\simeq\varepsilon(z)$, then we can write
\begin{align}
  & \left(\frac{B^2}{8\pi}\right)_z = \frac{\partial}{\partial z} \frac{B^2}{8\pi}\simeq \frac{B_{x0}^2-B_{xm}^2}{8\pi \Delta z}\\ 
  & (\nabla P_\perp)_z = \frac{\partial P_\perp}{\partial z}\simeq  \frac{P_{0}+P_{i0}(\sqrt{2\beta_{i0}}+\bar{B}_{xm})}{\sqrt{2\beta_{i0}}+1}\frac{1-\bar{B}_{xm}}{\Delta z}\\
  & \left[\nabla\cdot\left(\varepsilon\frac{\vect{B}\vect{B}}{4\pi}\right)\right]_z =\frac{\partial}{\partial x}\left(\frac{\varepsilon B_xB_z}{4\pi} \right)+\frac{\partial}{\partial z}\left(\frac{\varepsilon B_zB_z}{4\pi} \right)\simeq \frac{\varepsilon_1}{4\pi}\frac{B_{x0}+B_{xm}}{2}\frac{2B_{zh}}{\Delta x},
\end{align}
where we have used $B_z=0$ along the inflow symmetry line in Fig.~\ref{fig:schematic}(a), and
$B_{zh}$ is evaluated at the separatrix. The anisotropy parameter at point 1 is
\begin{align}
  \varepsilon_1 &
  = 1 + \frac{4\pi P_{i0}(\bar{B}_{x1}-1)}{B_{x1}^2}\left(1-\frac{1-\bar{B}_{x1}}{\sqrt{2\beta_{i0}}+1}\right)
  = 1 + \frac{\beta_{i0}}{2}\left(\frac{1}{\bar{B}_{x1}} - \frac{1}{\bar{B}_{x1}^2}\right)\left(1-\frac{1-\bar{B}_{x1}}{\sqrt{2\beta_{i0}}+1}\right),
\end{align}
where
\begin{equation}
  \bar{B}_{x1}=\frac{B_{x1}}{B_{x0}}
  = \frac{1+\bar{B}_{xm}}{2}.
\end{equation}
Since $\Delta z/\Delta x\simeq B_{zh}/B_{xh} \simeq B_{zh}/B_{x1}\simeq B_{zm}/B_{xm}$, then
\begin{equation}
  B_{zh}\simeq \frac{B_{zm}}{B_{xm}}B_{xh}\simeq
  \frac{B_{zm}}{B_{xm}}\frac{B_{x0}+B_{xm}}{2}\simeq \frac{\Delta z}{\Delta x}\frac{B_{x0}+B_{xm}}{2}.
\end{equation}
The inflow force balance becomes
\begin{align}
  1 - \bar{B}_{xm}^2 + \frac{\beta_{0}+\beta_{i0}(\sqrt{2\beta_{i0}}+\bar{B}_{xm})}{\sqrt{2\beta_{i0}}+1}\left(1-\bar{B}_{xm}\right)
  = \varepsilon_1\left(\frac{\Delta z}{\Delta x}\right)^2
  \left(1+\bar{B}_{xm}\right)^2,
\end{align}
where $\beta_{0}\equiv8\pi P_{0}/B_{x0}^2$.
We can numerically solve this equation to get $\bar{B}_{xm}$ for a given slope $\Delta z/\Delta x$ (i.e, the opening angle $\theta\equiv\tan^{-1}(\Delta z/\Delta x)$) and plasma $\beta_0$.

\label{app:heating}
\section{Outflow force balance}
\label{app:outflow}
According the momentum equation (Eq.~(\ref{equ:mom})),
the force balance equation that describes the outflow is
\begin{equation}
  nm_i\vect{V}\cdot\nabla\vect{V} + \nabla\frac{B^2}{8\pi} + \nabla P_\perp
  \simeq\nabla\cdot\left(\varepsilon \frac{\vect{B}\vect{B}}{4\pi}\right).
\end{equation}
At point 2 along the outflow direction (see Fig.~\ref{fig:schematic}(b)),
\begin{align}
  \frac{n_2m_i}{2}\frac{\partial V_x^2}{\partial x} + \frac{\partial}{\partial x}
  \frac{B^2}{8\pi} + \frac{\partial P_\perp}{\partial x} & =
  \frac{\partial}{\partial x}\left(\frac{\varepsilon B_xB_x}{4\pi}\right) +
  \frac{\partial}{\partial z}\left(\frac{\varepsilon B_zB_x}{4\pi}\right),
\end{align}
where $n_2$ is the plasma density at point 2 in Fig.~\ref{fig:schematic}(b). We again approximate $B_x\simeq B_x(z)$, $B_z\simeq B_z(x)$, and $\varepsilon\simeq\varepsilon(z)$ to simplify the problem. Since the magnetic field is primarily along the $z$-direction near the midplane ($z=0$), $\partial P_\perp/\partial x\simeq\partial P_{xx}/\partial x$.

When ions are reflected by the outflow, they gain energy through the Fermi mechanism. Only for ions that move toward the X-point, or those that move away from the X-point with a speed lower than the outflow speed $V$ will be reflected. As shown in Fig.~\ref{fig:schematic}(c), particle velocity changes from $v_x$ to $2V-v_x$ during the reflection. These particles will interpenetrate with incoming particles that have not been reflected yet and have $v_x$ from $-\infty$ to $V$; i.e., particles with $v_x>V$ have escaped from the diffusion region (Fig.~\ref{fig:schematic}(d)). The average velocity of these two populations is the outflow velocity $V$, as expected. This mixture results in a plasma density depending on $V$
\begin{align}
  \label{equ:rho}
  n(V) = 2\int_{-\infty}^V f(v_x)dv_x =
  n_0 + n_0\erf\left(\sqrt{\frac{m_i}{2kT_0}}V\right),
\end{align}
where we use $f(v_x)=n_0\sqrt{m_i/(2\pi kT_0)}\exp(-m_iv_x^2/(2kT_0))$, a
one-dimensional Maxwellian distribution with a temperature $T_0$ and a density
$n_0$. Here $k$ is the Boltzmann constant. Then, the $x$-component of the diagonal part of the ion pressure tensor is
\begin{align}
  P_{ixx} & = 2m_i\int_{-\infty}^V (V-v_x)^2f(v_x)dv_x \nonumber \\
  & = n_0m_i\left[\left(V^2 + \frac{kT_0}{m_i}\right)
  \left(1+\erf\left(\sqrt{\frac{m_i}{2kT_0}}V\right)\right)
  + V\sqrt{\frac{2kT_0}{\pi m_i}}e^{-m_iV^2/2kT_0}\right].
\end{align}
Since the pressure $P_{ixx}$ right at the x-line depleted by the inflowing plasma to near the background value, the increase of the ion pressure along the outflow can be modeled as
\begin{align}
  \label{equ:dpxx}
  \Delta P_{ixx} = P_{ixx} - n_0kT_0
  & = n_0m_i\left[V^2 + \left(V^2 + \frac{kT_0}{m_i}\right)
  \erf\left(\sqrt{\frac{m_i}{2kT_0}}V\right)
  + V\sqrt{\frac{2kT_0}{\pi m_i}}e^{-m_iV^2/2kT_0}\right].
\end{align}
Inspired by in-situ observations (and our simulations) that show a stronger ion heating than electron heating in reconnection exhausts~\citep{Phan2013Electron,Phan2014Ion}, we ignore the electron heating in the following analysis. The outflow force balance equation becomes
\begin{align}
  \frac{n_0+n_m}{2}\frac{m_i V_\text{out,m}^2}{2L} + \frac{B_{zm}^2}{8\pi L} + \frac{\Delta P_{ixx,m}}{L}
  & = \frac{1}{4\pi}\frac{B_{zm}}{2}\frac{(\varepsilon_mB_{xm})/2}{\delta/2},
  \label{equ:balance_outflow_o}
\end{align}
where we approximate $n_2=(n_0+n_m)/2$, $n_m$ is evaluated from Eq.~(\ref{equ:rho}) using $V=V_\text{out,m}$, $\Delta P_{ixx,m}$ is evaluated from Eq.~(\ref{equ:dpxx}) using $V=V_\text{out,m}$, and $\varepsilon_m=\varepsilon (B_{xm})$ is the anisotropy parameter immediately upstream of the diffusion region,
\begin{equation}
  \varepsilon_m
  = 1 + \frac{\beta_{i0}}{2}\left(\frac{1}{\bar{B}_{xm}} - \frac{1}{\bar{B}_{xm}^2}\right)\left(1-\frac{1-\bar{B}_{xm}}{\sqrt{2\beta_{i0}}+1}\right).
\end{equation}
Dividing Eq.~(\ref{equ:balance_outflow_o}) by $n_0m_iv_{A0}^2$, we get the normalized equation
\begin{align}
  \frac{1}{4}\left[2+\erf\left(\frac{\bar{V}_\text{out,m}}{\sqrt{\beta_{i0}}}\right)\right]
  \bar{V}_\text{out,m}^2 + \Delta\bar{P}_{ixx,m} +
  \frac{\bar{B}_{xm}^2}{2}\left[\left(\frac{\Delta z}{\Delta x}\right)^2 - \varepsilon_m\right] = 0,
\end{align}
where we have used $B_{zm}/B_{xm}\simeq\delta/L\simeq\Delta z/\Delta x$, $\beta_{i0}=8\pi P_{i0}/B_{x0}^2$, $\bar{V}_\text{out,m}=V_\text{out,m}/v_{A0}$, $\bar{B}_{xm}=B_{xm}/B_{x0}$, $\bar{B}_{zm}=B_{zm}/B_{x0}$, and
\begin{align}
  \Delta\bar{P}_{ixx,m} = \frac{4\pi\Delta P_{ixx,m}}{B_{x0}^2}
  = \bar{V}_\text{out,m}^2 +
  \left(\bar{V}_\text{out,m}^2 + \frac{\beta_{i0}}{2}\right)
  \erf\left(\frac{\bar{V}_\text{out,m}}{\sqrt{\beta_{i0}}}\right) +
  \bar{V}_\text{out,m}\sqrt{\frac{\beta_{i0}}{\pi}}e^{-\bar{V}_\text{out,m}^2/\beta_{i0}}.
\end{align}

\bibliography{references}{}
\bibliographystyle{aasjournal}
\end{document}